\documentclass[reqno]{article}
\usepackage{amsmath,amssymb,latexsym}

\newcommand\rf[1]{(\ref{eq:#1})}
\newcommand\lab[1]{\label{eq:#1}}
\newcommand\nonu{\nonumber}
\newcommand\br{\begin{eqnarray}}
\newcommand\er{\end{eqnarray}}
\newcommand\be{\begin{equation}}
\newcommand\ee{\end{equation}}

\newcommand\foot[1]{\footnotemark\footnotetext{#1}}
\newcommand\lb{\lbrack}
\newcommand\rb{\rbrack}

\newcommand\llb{\left\lbrack}
\newcommand\rrb{\right\rbrack}

\renewcommand\({\left(}
\renewcommand\){\right)}
\renewcommand\v{\vert}                     %% vertical bars
\newcommand\bgv{\bigg\vert}              %%

\newcommand\bc{\begin{center}}
\newcommand\ec{\end{center}}

\newcommand\Tr{\mathop{\mathrm Tr}}                  % Tr - big trace
\newcommand\partder[2]{\frac{{\partial {#1}}}{{\partial {#2}}}}
\renewcommand\d{\delta}

\newcommand\vareps{\varepsilon}
\newcommand\g{\gamma}
\newcommand\G{\Gamma}

\newcommand\h{\frac{1}{2}}
\renewcommand\k{\kappa}
\renewcommand\l{\lambda}
\renewcommand\L{\Lambda}
\newcommand\m{\mu}
\newcommand\n{\nu}
\renewcommand\o{\over}

\newcommand\p{\phi}
\newcommand\vp{\varphi}
\renewcommand\P{\Phi}
\newcommand\pa{\partial}

\newcommand\pr{\prime}

\renewcommand\r{\rho}
\newcommand\s{\sigma}
\renewcommand\S{\Sigma}
\renewcommand\t{\tau}
\renewcommand\th{\theta}

\newcommand\twomat[4]{\left(\begin{array}{cc}  %%   2x2 matrix  %ESA
{#1} & {#2} \\ {#3} & {#4} \end{array} \right)}

\newcommand\cA{{\mathcal A}}

\newcommand\cF{{\mathcal F}}

\newcommand\cM{{\mathcal M}}

\newcommand{\ct}[1]{\cite{#1}}
\newcommand{\bib}[1]{\bibitem{#1}}
\newcommand\NPB[3]{(#2), \textsl{Nucl. Phys.} \textbf{B#1} #3}

\newcommand\PRD[3]{(#2), \textsl{Phys. Rev.} \textbf{D#1} #3}

\newcommand\PLB[3]{(#2), \textsl{Phys. Lett.} \textbf{#1B} #3}
\newcommand\CQG[3]{(#2), \textsl{Class. Quantum Grav.} \textbf{#1} #3}

\newcommand\IJMPA[3]{(#2), \textsl{Int. J. Mod. Phys.} \textbf{A#1} #3}

\begin{document}

\title{Novel Aspects in $p$-Brane Theories: \\
Weyl-Invariant Light-Like Branes}

\bigskip
\author{Eduardo Guendelman and Alexander Kaganovich\\
\small\it Department of Physics, Ben-Gurion University, Beer-Sheva, Israel  \\[-1.mm]
\small\it email: guendel@bgumail.bgu.ac.il, alexk@bgumail.bgu.ac.il \\
${}$ \\
Emil Nissimov and Svetlana Pacheva\\
\small\it Institute for Nuclear Research and Nuclear Energy,\\[-1.mm]
\small\it Bulgarian Academy of Sciences, Sofia, Bulgaria  \\[-1.mm]
\small\it email: nissimov@inrne.bas.bg, svetlana@inrne.bas.bg}
\date{ }
\maketitle
\bigskip

\begin{abstract}
We consider a novel class of Weyl-conformally invariant $p$-brane theories which
describe intrinsically light-like branes for any odd world-volume dimension,
hence the acronym \textsl{WILL}-branes (Weyl-Invariant Light-Like branes).
We discuss in some detail the properties of \textsl{WILL}-brane dynamics
which significantly differs from ordinary Nambu-Goto brane dynamics. We provide 
explicit solutions of \textsl{WILL}-membrane (i.e., $p=2$) equations of motion
in arbitrary $D=4$ spherically symmetric static gravitational backgrounds, as well
as in product spaces of interest in Kaluza-Klein context. In the first case we 
find that the \textsl{WILL}-membrane materializes the event horizon of the 
corresponding black hole solutions, thus providing an explicit dynamical 
realization of the membrane paradigm in black hole physics. In the second 
``Kaluza-Klein'' context we find solutions describing \textsl{WILL}-branes
wrapped around the internal (compact) dimensions and moving as a whole with
the speed of light in the non-compact (space-time) dimensions.

\end{abstract}   

%%%%%%%%%%%%%%%%%%%%%%%%%%%%%%%%%%%%%%%%%%%%%%%%%%%%%%%%%%%%%%%%%%%%%%%%%%%%%%%%
\section{Introduction}   

The idea of replacing the standard Riemannian integration measure
(Riemannian volume-form) with an alternative non-Riemannian volume-form or,
more generally, employing on equal footing both Riemannian and non-Riemannian 
volume-forms to construct new classes of models involving gravity, called 
\textsl{two-measure theories}, has been proposed few years ago \ct{TMT-basic}
and since then it is a subject of active research and developments
\ct{TMT-recent} (for related ideas, see \ct{Hehl-etal}).

Two-measure theories address various basic problems in cosmology and
particle physics, and provide plausible solutions for a broad array of issues, 
such as:
scale invariance and its dynamical breakdown; spontaneous generation of
dimensionfull fundamental scales;  
the cosmological constant problem;  
the problem of fermionic families; 
applications in modern brane-world scenarios.
For a detailed discussion we refer to the series of papers \ct{TMT-basic,TMT-recent}.

Subsequently, the idea of employing an alternative non-Riemannian integration
measure was applied systematically to string, $p$-brane and $Dp$-brane models
\ct{m-string} (for a background on string and brane theories, see 
refs.\ct{brane-string-rev}). The main feature of these new classes of modified
string/brane theories is the appearance of the pertinent string/brane
tension as an additional dynamical degree of freedom beyond the usual string/brane
physical degrees of freedom, instead of being introduced \textsl{ad hoc} as
a dimensionfull scale. The dynamical string/brane tension acquires the
physical meaning of a world-sheet electric field strength (in the string
case) or world-volume $p+1$-form field strength (in the $p$-brane case) and
obeys Maxwell (Yang-Mills) equations of motion or their higher-rank
antisymmetric tensor gauge field analogues, respectively. As a result of the
latter property the modified-measure string model with dynamical tension
yields a simple classical mechanism of ``color'' charge confinement. 

In the next section we proceed to our main task which is the study of a 
novel class (first proposed in our preceding work \ct{will-brane}) of 
$p$-brane theories which are Weyl-conformal invariant for any $p$~ and 
which describe intrinsically light-like branes for any odd $(p+1)$. Thus,
their dynamics significantly differs both from the standard Nambu-Goto (or
Dirac-Born-Infeld) branes as well as from their modified versions with
dynamical string/brane tensions \ct{m-string} mentioned above. 
% %%%%%%%%%%%%%%%%%%%%%%%%%%%%%%%%%%%%%%%%%%%%%%%%%%%%%%%%%%%%%%%%%%%%%%%%%%%%%%%%

\section{Weyl-Invariant $p$-Brane Theories}

\subsection{Standard Nambu-Goto Branes}
Let us first briefly recall the standard Polyakov-type formulation of the 
bosonic $p$-brane action:
\be
S = -{T\o 2}\int d^{p+1}\s\,\sqrt{-\g}\, \Bigl\lb 
\g^{ab}\pa_a X^\m \pa_b X^\n G_{\m\n}(X) - \L (p-1)\Bigr\rb \; .
\lab{stand-brane-action}
\ee
Here $\g_{ab}$ is the ordinary Riemannian metric on the $p+1$-dimensional
brane world-volume with $\g \equiv \det\v\v \g_{ab}\v\v$. The world-volume indices
$a,b=0,1,\ldots ,p$ ; 
~$G_{\m\n}$ denotes the Riemannian metric in the embedding space-time with 
space-time indices $\m,\n=0,1,\ldots ,D-1$.
$T$ is the given \textsl{ad hoc} brane tension; the constant
$\L$ can be absorbed by rescaling $T$ (see below Eq.\rf{stand-brane-action-NG}).
The equations of motion w.r.t. $\g^{ab}$ and $X^\m$ read:
\be
T_{ab} \equiv \( \pa_a X^\m \pa_b X^\n - 
\h \g_{ab} \g^{cd}\pa_c X^\m \pa_d X^\n \) G_{\m\n} + \g_{ab} {\L\o 2}(p-1)
= 0 \; ,
\lab{stand-brane-gamma-eqs}
\ee
\be
\pa_a \(\sqrt{-\g}\g^{ab}\pa_b X^\m\) +
\sqrt{-\g}\g^{ab}\pa_a X^\n \pa_b X^\l \G^\m_{\n\l} = 0  \; ,
\lab{stand-brane-X-eqs}
\ee
where:
\be
\G^\m_{\n\l}=\h G^{\m\k}\(\pa_\n G_{\k\l}+\pa_\l G_{\k\n}-\pa_\k G_{\n\l}\)
\lab{affine-conn}
\ee
is the affine connection for the external metric.

Eqs.\rf{stand-brane-gamma-eqs} when $p \neq 1$ imply:
\be
\L \g_{ab} = \pa_a X^\m \pa_b X^\n G_{\m\n} \; ,
\lab{stand-brane-metric-eqs}
\ee
which in turn allows to rewrite Eq.\rf{stand-brane-gamma-eqs} as:
\be
T_{ab} \equiv \( \pa_a X^\m \pa_b X^\n - 
{1\o {p+1}} \g_{ab} \g^{cd}\pa_c X^\m \pa_d X^\n \) G_{\m\n} = 0 \; .
\lab{stand-brane-gamma-eqs-Pol}
\ee
Furthermore, using \rf{stand-brane-metric-eqs} the Polyakov-type brane action 
\rf{stand-brane-action} becomes on-shell equivalent to the Nambu-Goto-type brane
action:
\be
S = - T \L^{-{{p-1}\o 2}} \int d^{p+1}\s\,
\sqrt{-\det\v\v \pa_a X^\m \pa_b X^\n G_{\m\n} \v\v} \; .
\lab{stand-brane-action-NG}
\ee

\subsection{Weyl-Invariant Branes: Action and Equations of Motion}

In ref.\ct{will-brane} we proposed the following novel $p$-brane actions:
\be
S = - \int d^{p+1}\s \,\P (\vp) 
\Bigl\lb \h \g^{ab} \pa_a X^{\m} \pa_b X^{\n} G_{\m\n}(X)
- \sqrt{F_{ab}(A) F_{cd}(A) \g^{ac}\g^{bd}}\Bigr\rb
\lab{WI-brane}
\ee
with $F_{ab}(A) = \pa_a A_b - \pa_b A_a$, and:
\be
\P (\vp) \equiv \frac{1}{(p+1)!} \vareps_{i_1\ldots i_{p+1}} 
\vareps^{a_1\ldots a_{p+1}} \pa_{a_1} \vp^{i_1}\ldots \pa_{a_{p+1}} \vp^{i_{p+1}}
\quad ,\quad i,j=1,\ldots,p+1 \; .
\lab{mod-measure-p}
\ee
Here $\g_{ab}$ and $G_{\m\n}$ have the same meaning as in \rf{stand-brane-action}.

Let us notice the following significant differences of \rf{WI-brane} w.r.t. 
the standard Nambu-Goto $p$-branes (in the Polyakov-like formulation) 
\rf{stand-brane-action}:
\begin{itemize}
\item
New non-Riemannian integration measure density (volume-form)
$\P (\vp)$ \rf{mod-measure-p} instead of the usual $\sqrt{-\g}$,
built entirely in terms of auxiliary world-sheet scalar fields $\vp^i$
independent of the Riemannian metric $\g_{ab}$.
\item
There is {\em no}~ ``cosmological-constant'' term ($(p-1)\sqrt{-\g}$) in 
\rf{WI-brane}.
\item
The action \rf{WI-brane} is manifestly Weyl-conformal invariant for {\em any} $p$;
here Weyl-\-conformal symmetry is given by Weyl rescaling of $\g_{ab}$ supplemented
with a special diffeomorphism in the target space of auxiliary $\vp$-fields:
\be
\g_{ab} \longrightarrow \g^{\pr}_{ab} = \rho\,\g_{ab}  \quad ,\quad
\vp^{i} \longrightarrow \vp^{\pr\, i} = \vp^{\pr\, i} (\vp) 
\;\; \mathrm{with} \;\; 
\det \Bigl\Vert \frac{\pa\vp^{\pr\, i}}{\pa\vp^j} \Bigr\Vert = \rho \; .
\lab{Weyl-conf}
\ee
\item
There are {\em no}~ \textsl{ad hoc}~ dimensionfull constants in \rf{WI-brane};
the {\em variable} brane tension $\chi \equiv \frac{\P (\vp)}{\sqrt{-\g}}$ 
is Weyl-\-conformal {\em gauge dependent}: $ \chi \to \rho^{\h(1-p)}\chi$.
\item
The action \rf{WI-brane} contains an additional world-volume gauge field $A_a$ in a 
``square-root'' Maxwell (Yang-Mills) Lagrangian\foot{``Square-root'' Maxwell 
(Yang-Mills) action in $D=4$ was originally introduced in \ct{Nielsen-Olesen} 
and later formulated in dual variables and generalized to ``square-root'' actions
of higher-rank antisymmetric tensor gauge fields in $D\geq 4$ in 
refs.\ct{Spallucci}; see also ref.\ct{Amer-Eduardo}.};
the latter can be straightforwardly generalized to the non-Abelian case: 
$\sqrt{-\Tr \( F_{ab}(A) F_{cd}(A)\) \g^{ac}\g^{bd}}$ with
$F_{ab}(A) = \pa_a A_b - \pa_b A_a + i \lb A_a, A_b\rb$ .
\item
The presence of the world-volume gauge field $A_a$ allows for natural (linear) 
optional couplings both to external world-volume as well as to space-time
``color'' charge currents in a Weyl-conformally invariant way (see 
Eq.\rf{WILL-membrane+A-a} below).
\item
The action \rf{WI-brane} describes {\em intrinsically light-like} $p$-branes for 
any odd $(p+1)$ (see Eq.\rf{LL-constraints} below).
\end{itemize}

The action \rf{WI-brane} yields the following equations of motion w.r.t. 
auxiliary scalars $\vp^i$:
\be
\h \g^{cd}\(\pa_c X \pa_d X\) - \sqrt{FF\g\g} = M \; \Bigl( = \mathrm{const}\Bigr)
\; ,
\lab{phi-eqs}
\ee
with the short-hand notations: 
\be
\(\pa_a X \pa_b X\) \equiv \pa_a X^\m \pa_b X^\n G_{\m\n}\quad ,\quad
\sqrt{FF\g\g} \equiv \sqrt{F_{ab} F_{cd} \g^{ac}\g^{bd}} \; .
\lab{short-hand}
\ee
The equations of motion w.r.t. $\g^{ab}$ are:
\be
\h\(\pa_a X \pa_b X\) + \frac{F_{ac}\g^{cd} F_{db}}{\sqrt{FF\g\g}} = 0 \; ,
\lab{gamma-eqs}
\ee
which upon taking the trace imply $M=0$ in Eq.\rf{phi-eqs}.

Further we obtain the following equations of motion w.r.t. world-volume gauge 
field $A_a$ and w.r.t. brane embedding coordinates $X^\m$, respectively:
\be
\pa_b \(\frac{F_{cd}\g^{ac}\g^{bd}}{\sqrt{FF\g\g}} \P (\vp)\) = 0 \; ,
\lab{A-eqs}
\ee
\be
\pa_a \(\P (\vp) \g^{ab}\pa_b X^\m\) +
\P (\vp) \g^{ab}\pa_a X^\n \pa_b X^\l \G^\m_{\n\l} = 0 \; ,
\lab{X-eqs}
\ee
where $\G^\m_{\n\l}$ is the same as in \rf{affine-conn}.

\subsection{Light-Like Branes}

Now, let us consider the $\g^{ab}$-equations of motion \rf{gamma-eqs}. Since
$F_{ab}$ is an anti-symmetric $(p+1)\times (p+1)$ matrix, it is therefore
{\em not invertible} in any odd $(p+1)$, \textsl{i.e.} $F_{ab}$ has at least one 
zero-eigenvalue vector $V^a$ ($F_{ab}V^b = 0$). Thus, for any odd $(p+1)$ the 
induced metric:
\be
g_{ab} \equiv \(\pa_a X \pa_b X\) \equiv \pa_a X^\m \pa_b X^\n G_{\m\n}
\lab{ind-metric}
\ee
on the world-volume of the Weyl-invariant brane \rf{WI-brane} is {\em singular} 
as {\em opposed} to the ordinary Nambu-Goto brane where the induced metric
is proportional to the intrinsic Riemannian world-volume metric 
(cf. Eq.\rf{stand-brane-metric-eqs}). In other words:
\be
\(\pa_a X \pa_b X\) V^b = 0 \quad ,\quad \mathrm{i.e.}\;\;
\(\pa_V X \pa_V X\) = 0 \;\; ,\;\; \(\pa_{\perp} X \pa_V X\) = 0 \; ,
\lab{LL-constraints}
\ee
where $\pa_V \equiv V^a \pa_a$ and $\pa_{\perp}$ are derivates along the
tangent vectors in the complement of the tangent vector field $V^a$. 

The constraints \rf{LL-constraints} imply the following important conclusion:
every point on the (fixed-time) world-surface of the Weyl-invariant $p$-brane 
\rf{WI-brane} (for odd $(p+1)$) moves in orthogonal direction w.r.t. itself with 
the speed of light in a time-evolution along the zero-eigenvalue vector-field 
$V^a$ of the world-volume electromagnetic field-strength $F_{ab}$. 
Therefore, we will call \rf{WI-brane} (for odd $(p+1)$) by the acronym 
{\em WILL-brane} (Weyl-Invariant Light-Like-brane) model.
  
\subsection{Dual Formulation of {\em WILL}-Branes}

The $A_a$-equations of motion \rf{A-eqs} can be solved in terms of $(p-2)$-form gauge 
potentials $\L_{a_1\ldots a_{p-2}}$ dual w.r.t. $A_a$. The respective
field-strengths are related as follows:
\be
F_{ab}(A)= -\frac{1}{\chi}\,\frac{\sqrt{-\g}\,\vareps_{abc_1\ldots c_{p-1}}}{2(p-1)}
\g^{c_1 d_1}\ldots \g^{c_{p-1} d_{p-1}} 
\, F_{d_1\ldots d_{p-1}}(\L) \,\g^{cd} \(\pa_c X \pa_d X\) \; ,
\lab{dual-strength-rel}
\ee
where:
\be
F_{a_1\ldots a_{p-1}}(\L) = (p-1) \pa_{[a_1} \L_{a_2\ldots a_{p-1}]}
\lab{dual-strength}
\ee
is the $(p-1)$-form dual field-strength, and
$\chi \equiv \frac{\P (\vp)}{\sqrt{-\g}}$ is the variable brane tension,
which we find to be explicitly expressed in terms of the dual field-strength:
\be
\chi^2 \equiv \chi^2 (\g,\L) =
- \frac{2}{(p-1)^2}\, \g^{a_1 b_1}\ldots \g^{a_{p-1} b_{p-1}}
F_{a_1\ldots a_{p-1}}(\L) F_{b_1\ldots b_{p-1}}(\L) \; .
\lab{chi-2}
\ee
Now, the Biancchi identities for $A_a$ turn into dynamical equations of
motion for the dual $(p-2)$-form gauge potentials $\L_{a_1\ldots a_{p-2}}$:
\be
\pa_a \(\frac{\sqrt{-\g}}{\chi (\g,\L)} 
\g^{ab}\g^{a_1 b_1}\ldots \g^{a_{p-2} b_{p-2}} F_{b b_1\ldots b_{p-2}} (\L)\,
\g^{cd}\(\pa_c X \pa_d X\)\) = 0
\lab{A-dual-eqs}
\ee

All equations of motion \rf{gamma-eqs},\rf{X-eqs} and \rf{A-dual-eqs} can be 
equivalently derived from the following {\em dual} {\em WILL}-brane action:
\be
S_{\mathrm{dual}} = - \h \int d^{p+1}\s\, \chi (\g,\L) \sqrt{-\g}
\g^{ab}\pa_a X^\m \pa_b X^\n G_{\m\n}
\lab{WI-brane-dual}
\ee
with $\chi (\g,\L)$ given in \rf{chi-2} above.
  
\section{The {\em WILL}-Membrane}

The {\em WILL}-membrane dual action (particular case of \rf{WI-brane-dual} for
$p=2$) reads:
\br
S_{\mathrm{dual}} = - \h \int d^3\s\, \chi (\g,u)\,\sqrt{-\g}
\g^{ab}\(\pa_a X \pa_b X\) \; ,
\lab{WILL-membrane} \\
\chi (\g,u) \equiv \sqrt{-2\g^{cd}\pa_c u \pa_d u} \; ,
\lab{chi-1}
\er
where $u$ is the dual ``gauge'' potential w.r.t. $A_a$:
\be
F_{ab}(A) = - \frac{1}{2\chi (\g,u)} \sqrt{-\g} \vareps_{abc} \g^{cd}\pa_d u\,
\g^{ef}\!\(\pa_e X \pa_f X\)  \; .
\lab{dual-strenght-rel-3}
\ee
$S_{\mathrm{dual}}$ is manifestly Weyl-invariant (under $\g_{ab} \to \rho\g_{ab}$).

The equations of motion w.r.t. $\g^{ab}$, $u$ (or $A_a$), and $X^\m$ read
accordingly:
\be
\(\pa_a X \pa_b X\) + \h \g^{cd}\(\pa_c X \pa_d X\) 
\(\frac{\pa_a u \pa_b u }{\g^{ef} \pa_e u \pa_f u} - \g_{ab}\) = 0 \; ,
\lab{gamma-eqs-3}
\ee
\be
\pa_a \(\,\frac{\sqrt{-\g}\g^{ab}\pa_b u}{\chi (\g,u)}\,
\g^{cd}\(\pa_c X \pa_d X\)\,\) = 0 \; ,
\lab{u-eqs}
\ee
\be
\pa_a \(\chi (\g,u)\,\sqrt{-\g} \g^{ab}\pa_b X^\m \) +
\chi (\g,u)\,\sqrt{-\g} \g^{ab}\pa_a X^\n \pa_b X^\l \G^\m_{\n\l} = 0 \; .
\lab{X-eqs-3}
\ee
The first equation above shows that the induced metric 
$g_{ab} \equiv \(\pa_a X \pa_b X\)$ has zero-mode eigenvector $V^a =\g^{ab}\pa_b u$.

The invariance under world-volume reparametrizations allows to introduce the
following standard (synchronous) gauge-fixing conditions:
\be
\g^{0i} = 0 \;\; (i=1,2) \quad ,\quad \g^{00} = -1 \; .
\lab{gauge-fix}
\ee
% There remains a residual $\t \equiv \s^0$-independent reparametrization invariance
% allowing for further con\-for\-mally-flat gauge-fixing of the space-like part of 
% $\g_{ab}$:
% \be
% \g_{ij} = a (\t,\s^1,\s^2) {\wti \g}_{ij} (\s^1,\s^2) \; ,
% \lab{conf-flat}
% \ee
% with ${\wti \g}_{ij}$ a standard reference $2D$ metric on the membrane
% surface ($i,j=1,2$).

In spite of the high non-linearity of Eq.\rf{u-eqs} for the dual ``gauge
potential'' $u$, we can easily find solutions by using the following ansatz:
\be
u (\t,\s^1,\s^2) = \frac{T_0}{\sqrt{2}}\t  \; ,
\lab{u-ansatz}
\ee
where $T_0$ is an arbitrary integration constant with the dimension of membrane
tension. In particular:
\be
\chi \equiv \sqrt{-2\g^{ab}\pa_a u \pa_b u} = T_0 
\lab{chi-0}
\ee
The ansatz \rf{u-ansatz} means that we take $\t\equiv\s^0$ to be evolution
parameter along the zero-eigenvalue vector-field of the induced metric on the brane 
($V^a = \g^{ab}\pa_b u = \mathrm{const}\,(1,0,0)$).

% \textbf{Gauge-Fixed Equations of Motion}  
  
With the gauge choice for $\g_{ab}$ \rf{gauge-fix} the equations of motion w.r.t.
$\g^{ab}$ \rf{gamma-eqs-3} (which are in fact constraints) become
(recall $\(\pa_a X \pa_b X\) \equiv \pa_a X^\m \pa_b X^\n G_{\m\n}$):
\be
\(\pa_0 X \pa_0 X\) = 0 \quad ,\quad \(\pa_0 X \pa_i X\) = 0  \; ,
\lab{constr-0}
\ee
\be
\(\pa_i X\pa_j X\) - \h \g_{ij} \g^{kl}\(\pa_k X\pa_l X\) = 0  \; ,
\lab{constr-vir}
\ee
Note that Eqs.\rf{constr-vir} look exactly like the classical (Virasoro) 
constraints for an Euclidean string theory with world-sheet parameters 
$(\s^1,\s^2)$.
%%%%%%%%%%%%%%%%

The gauge choice for \rf{gauge-fix} together with the ansatz 
\rf{u-ansatz}, as well as taking into account \rf{constr-0}, bring the
the equations of motion w.r.t. $u$ to the form:
\be
\pa_0 \(\sqrt{\g_{(2)}} \g^{kl}\(\pa_k X\pa_l X\)\) = 0  \; ,
\lab{u-eqs-fix}
\ee
where $\g_{(2)} = \det\Vert \g_{ij}\Vert$ ($i,j,k,l=1,2$). Eq.\rf{u-eqs-fix}
is the only remnant from the original $A_a$-equations of motion \rf{A-eqs}.
%%%%%%%%%%%%%%%%

Accordingly, the $X^\m$-equations of motion now read:
\be
\Box^{(3)} X^\m + \( - \pa_0 X^\n \pa_0 X^\l + 
\g^{kl} \pa_k X^\n \pa_l X^\l \) \G^{\m}_{\n\l} = 0  \; ,
\lab{X-eqs-3-fix}
\ee
where:
\be
\Box^{(3)} \equiv 
- \frac{1}{\sqrt{\g^{(2)}}} \pa_0 \(\sqrt{\g^{(2)}} \pa_0 \) + 
\frac{1}{\sqrt{\g^{(2)}}}\pa_i \(\sqrt{\g^{(2)}} \g^{ij} \pa_j \)   \; .
\lab{box-3}
\ee
We recall that everywhere in Eqs.\rf{constr-0}--\rf{box-3} the space-like
part of the internal membrane metric $\g_{ij}$ is of the form \rf{conf-flat}.
%%%%%%%%%%%%%%%%%%%%%%%%%%%%%%%%%%%%%%%%%%%%%%%%%%%%%%%%%%%%%%%%%%%%%%%%%%%

\section{{\em WILL}-Memrane Solutions in Non-Trivial Gravitational Backgrounds}    

\subsection{Example: {\em WILL}-Membrane in Spherically-Symmetric Static 
Backgrounds}

Let us consider a general spherically-symmetric static gravitational
background in $D=4$ embedding space-time:
\be
(ds)^2 = - A(r)(dt)^2 + B(r)(dr)^2 + 
r^2 \lb (d\th)^2 + \sin^2 (\th)\,(d\p)^2\rb \; .
\lab{spherical-symm-metric}
\ee
Specifically we have:
\be
A(r) = B^{-1}(r) = 1 - \frac{2GM}{r}
\lab{schwarzschild}
\ee
for Schwarzschild black hole,
\be
A(r) = B^{-1}(r) = 1 - \frac{2GM}{r} + \frac{Q^2}{r^2}
\lab{R-N}
\ee
for Reissner-Nordstr\"{o}m black hole,
\be
A(r) = B^{-1}(r) = 1 - \k r^2
\lab{AdS}
\ee
for (anti-) de Sitter space, \textsl{etc.}.

To find solutions of the equations of motion (and constraints) 
\rf{constr-0}--\rf{box-3} we will use the following ansatz:
\be
X^0 \equiv t = \t \quad,\quad X^1 \equiv r = r(\t,\s^1,\s^2) \quad, \quad
% X^2 \equiv \th = \th (\s^1,\s^2) \quad ,\quad  X^3 \equiv \p = \p (\s^1,\s^2) \; ,
X^2 \equiv \th = \s^1 \quad ,\quad  X^3 \equiv \p = \s^2 \; ;
\lab{ansatz-spherical-symm}
\ee
\be
\Vert\g_{ij}\Vert = a(\t)\,\twomat{1}{0}{0}{\sin^2 (\s^1)}
\lab{conf-flat}
\ee
In other words, we assume that the underlying \textsl{WILL}-membrane has
spherical topology of its fixed-time world-surface.

From Eqs.\rf{constr-0} taking into account \rf{spherical-symm-metric} we
obtain:
\be
\partder{}{\t}r = \pm A(r) \quad ,\quad \partder{}{\s^i}r = 0 \; .
\lab{sol-spherical-1}
\ee
From Eq.\rf{u-eqs-fix} we get $\partder{}{\t}r = 0$ which upon combining with 
\rf{sol-spherical-1} gives:
\be
r = r_0 \equiv \mathrm{const} \;\;,\;\; \mathrm{where} \quad A(r_0)=0  \; .
\lab{sol-spherical-2}
\ee
The $X^0$-equation of motion (Eq.\rf{X-eqs-3-fix} for $\m=0$) implies for the 
intrinsic {\em WILL}-membrane metric:
\be
\Vert\g_{ij}\Vert = c_0 \, e^{\mp \t/r_0}\,\twomat{1}{0}{0}{\sin^2 (\s^1)} \; ,
\lab{sol-spherical-3}
\ee
where $c_0$ is an arbitrary integration constant.

From \rf{sol-spherical-2} we conclude that the {\em WILL}-membrane with spherical 
topology (and with exponentially blowing-up/deflating radius w.r.t. internal metric)
``sits'' on (materializes) the event horizon of the pertinent black hole in
$D=4$ embedding space-time.
%%%%%%%%%%%%%%%%%%%%%%%%%%%%%%%%%%%%%%%%%%%%%%%%%%%%%%%%%%%%%%%%%%%%%%%%%%%%%%%%

\subsection{Example: {\em WILL}-membrane in Product-Space Backgrounds}

Here we consider {\em WILL}-membrane 
% with arbitrary $(p+1)=2k+1$ (odd-)dimensional world-volumes 
moving in a general product-space $D=(d+2)$-dimensional gravitational
background $\cM^d \times \S^{2}$ with coordinates $(x^\m, y^m)$ ($\m =
0,1,\ldots ,d-1$, $m = 1,2$) and Riemannian metric
$(ds)^2 = f(y) g_{\m\n}(x) dx^\m dx^\n + g_{mn}(y) dy^m dy^n$.

We assume that the {\em WILL}-brane wraps around the ``internal''
space $\S^{2}$ and use the following ansatz (recall $\t \equiv \s^0$):
\be
X^\m = X^\m (\t) \quad , \quad Y^m = \s^m \quad , \quad
\g_{mn} = a(\t)\, g_{mn}(\s^1,\s^2)
\lab{ansatz-product-space}
\ee
% (recall also Eq.\rf{conf-flat}).
Then the equations of motion and constraints \rf{constr-0}--\rf{box-3}
reduce to:
\be
\pa_\t X^\m \pa_\t X^\n g_{\m\n}(X) = 0 \quad ,\quad
\frac{1}{a(\t)}\,\pa_\t \Bigl( a(\t) \pa_\t X^\m \Bigr) + 
\pa_\t X^\n \pa_\t X^\l\, \G^\m_{\n\l} = 0 
\lab{eff-massless}
\ee
where $a(\t)$ is the conformal factor of the space-like part of the internal
membrane metric (last Eq.\rf{ansatz-product-space}). 
Eqs.\rf{eff-massless} are of the same form
as the equations of motion for a massless point-particle with a world-line
``einbein'' $e = a^{-1}$ moving in $\cM^d$. In other words, the simple solution 
above describes a membrane living in the extra ``internal'' dimensions and
moving as a whole with the speed of light in ``ordinary'' space-time.

Notice that although the {\em WILL}-brane is wrapping the extra dimensions in a
topologically non-trivial way (cf. second Eq.\rf{ansatz-product-space}), 
its modes remain {\em massless} from the
projected $d$-dimensional space-time point of view. This is a highly non-trivial
result since we have here particles (membrane modes), which aquire
in this way non-zero quantum numbers, while at the same time remaing
massless. In contrast, one should recall that in ordinary Kaluza-Klein
theory (for a review, see \ct{K-K-review}), 
non-trivial dependence on the extra dimensions is possible for
point particles or even standard strings and branes only at a very high
energy cost (either by momentum modes or winding modes), which implies a
very high mass from the projected $D=4$ space-time point of view. % \ct{Eduardo-91}.
%%%%%%%%%%%%%%%%%%%%%%%%%%%%%%%%%%%%%%%%%%%%%%%%%%%%%%%%%%%%%%%%%%%%%%%%%%%%%%%%

\subsection{Example: WILL-Membrane in a PP-Wave Background}
As a final non-trivial example let us consider \textsl{WILL}-membrane
dynamics in external plane-polarized gravitational wave (\textsl{pp-wave}) 
background:
\be
(ds)^2 = - dx^{+} dx^{-} - F(x^{+},x^I)\, (dx^{+})^2 + dx^I dx^I \; ,
\lab{pp-wave}
\ee
and employ in \rf{constr-0}--\rf{box-3} the following natural ansatz for $X^\m$ 
(here $\s^0 \equiv \t$; $I=1,\ldots,D-2$) :
\be
X^{-} = \t \quad, \quad 
X^{+}=X^{+}(\t,\s^1,\s^2) \quad, \quad X^I = X^I (\s^1,\s^2) \; .
\lab{ansatz-pp-wave}
\ee
The non-zero affine connection symbols for the pp-wave metric \rf{pp-wave}
are: $\G^{-}_{++}=\pa_{+}F$, $\G^{-}_{+I}=\pa_{I}F$, $\G^{I}_{++}=\h\pa_{I}F$.

It is straightforward to show that the solution does not depend on the form of 
the pp-wave front $F(x^{+},x^I)$ and reads:
\be
X^{+}=X^{+}_0 = \mathrm{const} \quad ,\quad 
\g_{ij} = \t\!-\!\mathrm{independent}\; ;
\lab{pp-wave-sol-1}
\ee
\be
\pa_i X^I \pa_j X^I - 
\h \g_{ij} \g^{kl} \pa_k X^I \pa_l X^I = 0 \quad ,\quad
\pa_i \(\sqrt{\g^{(2)}} \g^{ij} \pa_j X^I \) = 0 
\lab{pp-wave-sol-2}
\ee
where the latter equations describe a string embedded in the transverse 
$(D-2)$-dimensional flat Euclidean space.
%%%%%%%%%%%%%%%%%%%%%%%%%%%%%%%%%%%%%%%%%%%%%%%%%%%%%%%%%%%%%%%%%%%%%%%%%%%%%%%%

\section{{\em WILL}-Membrane as a Source for Gravity and Electromagnetism}

In this section we shall consider the Einstein-Maxwell system coupled to
an electrically charged {\em WILL}-membrane, \textsl{i.e.}, we shall take
into account the back-reaction of the {\em WILL}-membrane serving as a
material and electrically charged source for gravity and electromagnetism.
The relevant action reads:
\be
S = \int\!\! d^4 x\,\sqrt{-G}\,\llb \frac{R}{16\pi G_N}
- \frac{1}{4} \cF_{\m\n}(\cA) \cF_{\k\l}(\cA) G^{\m\k} G^{\n\l}\rrb
+ S_{\mathrm{WILL-brane}}  \; ,
\lab{E-M-WILL} 
\ee
where $\cF_{\m\n}(\cA) = \pa_\m \cA_\n - \pa_\n \cA_\m$, 
% $\cA_\m (x)$ being the space-time Maxwell gauge potential, 
and:
\br
S_{\mathrm{WILL-brane}} = - \int\!\! d^3\s \,\P (\vp) 
\Biggl\lb \h \g^{ab} \pa_a X^{\m} \pa_b X^{\n} G_{\m\n}
- \sqrt{F_{ab} F_{cd} \g^{ac}\g^{bd}}\,\Biggr\rb
- q\int\!\! d^3\s \, \vareps^{abc} \cA_\m \pa_a X^\m F_{bc}  \; .
\lab{WILL-membrane+A-a}
\er
Note the appearance of a natural Weyl-conformal invariant coupling of the
\textsl{WILL}-brane to the external space-time electromagnetic field
$\cA_\m$ -- the last Chern-Simmons-like term in \rf{WILL-membrane+A-a}.
The latter is a special case of a class of Chern-Simmons-like couplings of 
extended objects to external electromagnetic fields proposed in 
ref.\ct{Aaron-Eduardo}.

The Einstein-Maxwell equations of motion are of the standard form:
\be
R_{\m\n} - \h G_{\m\n} R = 8\pi G_N \( T^{(EM)}_{\m\n} + T^{(brane)}_{\m\n}\)\; ,
\lab{Einstein-eqs}
\ee
\be
\pa_\n \(\sqrt{-G}G^{\m\k}G^{\n\l} \cF_{\k\l}\) + j^\m = 0 \; ,
\lab{Maxwell-eqs}
\ee
where:
\be
T^{(EM)}_{\m\n} \equiv \cF_{\m\k}\cF_{\n\l} G^{\k\l} - G_{\m\n}\frac{1}{4}
\cF_{\r\k}\cF_{\s\l} G^{\r\s}G^{\k\l} \; ,
\lab{T-EM}
\ee
\be
T^{(brane)}_{\m\n} \equiv - G_{\m\k}G_{\n\l}
\int\!\! d^3 \s\, \frac{\d^{(4)}\Bigl(x-X(\s)\Bigr)}{\sqrt{-G}}\,
% \chi\,\sqrt{-\g} \g^{ab}\pa_a X^\k \pa_b X^\l \; ,
\P (\vp) \g^{ab}\pa_a X^\k \pa_b X^\l \; ,
\lab{T-brane}
\ee
% (recall $\chi \equiv \sqrt{-2\g^{cd}\(\pa_c u - q \cA_c\)
% \(\pa_d u - q \cA_d\)}$,  $\cA_a \equiv \cA_\m \pa_a X^\m$),
\be
j^\m \equiv q \int\!\! d^3 \s\,\d^{(4)}\Bigl(x-X(\s)\Bigr)
\vareps^{abc} F_{bc} \pa_a X^\m \; .
\lab{brane-EM-current}
\ee

For the \textsl{WILL}-membrane subsystem we can use instead of the action 
\rf{WILL-membrane+A-a} its dual one (similar to the simpler case Eq.\rf{WI-brane}
versus Eq.\rf{WILL-membrane}):
\be
S^{\mathrm{dual}}_{\mathrm{WILL-brane}}
= - \h \int d^3\s\, \chi (\g,u,\cA)\,\sqrt{-\g} \g^{ab}\(\pa_a X \pa_b X\)  \; ,
\lab{WILL-membrane+A-dual}
\ee
where the variable brane tension $\chi \equiv \frac{\P (\vp)}{\sqrt{-\g}}$ 
is given by:
\be
\chi (\g,u,\cA) \equiv \sqrt{-2\g^{cd}\(\pa_c u - q \cA_c\)
\(\pa_d u - q \cA_d\)} \quad ,\;\; \cA_a \equiv \cA_\m \pa_a X^\m \; .
\lab{tension+A}
\ee
Here $u$ is the dual ``gauge'' potential w.r.t. $A_a$ and the corresponding
field-strength and dual field-strength are related as
(cf. Eq.\rf{dual-strenght-rel-3}) :
\be
F_{ab}(A) = - \frac{1}{2\chi (\g,u,\cA)} \sqrt{-\g} \vareps_{abc} \g^{cd}
\(\pa_d u - q\cA_d \)\,\g^{ef}\!\(\pa_e X \pa_f X\)  \; .
\lab{dual-strenght-rel-3-A}
\ee
% The extended {\em WILL}-membrane model in the dual formulation 
% \rf{WILL-membrane+A-dual} is likewise manifestly Weyl-invariant 
% (under $\g_{ab} \to \rho\g_{ab}$).

The corresponding equations of motion w.r.t. $\g^{ab}$, $u$ (or $A_a$), and $X^\m$
read accordingly:
\be
\(\pa_a X \pa_b X\) + \h \g^{cd}\(\pa_c X \pa_d X\)
\(\frac{\(\pa_a u - q\cA_a\)\(\pa_b u -q\cA_b\) }
{\g^{ef} \(\pa_e u - q\cA_e\)\(\pa_f u - q\cA_f\)} - \g_{ab}\) = 0  \; ;
\lab{gamma-eqs+A}
\ee
\be
\pa_a \(\,\frac{\sqrt{-\g}\g^{ab}\(\pa_b u - q\cA_b\)}{\chi (\g,u,\cA)}\,
\g^{cd}\(\pa_c X \pa_d X\)\,\) = 0  \; ;
\lab{u-eqs+A}
\ee
\br
\pa_a \(\chi (\g,u,\cA)\,\sqrt{-\g} \g^{ab}\pa_b X^\m \) +
\chi (\g,u,\cA)\,\sqrt{-\g} \g^{ab}\pa_a X^\n \pa_b X^\l \G^\m_{\n\l} 
\nonu \\
- \; q \vareps^{abc} F_{bc} \pa_a X^\n 
\(\pa_\l \cA_\n - \pa_\n \cA_\l\)\, G^{\l\m} = 0  \; . \phantom{aaaaaaaa}
\lab{X-eqs+A}
\er

Following steps similar to the ones in the previous section we obtain the following
self-consistent spherically symmetric stationary solution for the full coupled
Einstein-Maxwell-{\em WILL}-membrane system \rf{E-M-WILL}. 
For the Einstein subsystem we have a solution:
\be
(ds)^2 = - A(r)(dt)^2 + A^{-1}(dr)^2 + r^2 \lb (d\th)^2 + \sin^2 (\th)\,(d\p)^2\rb
\; ,
\lab{spherical-symm-metric-b}
\ee
consisting of two different black holes with a {\em common} event horizon:
\begin{itemize}
\item    
Schwarzschild black hole inside the horizon:
\be
A(r)\equiv A_{-}(r) = 1 - \frac{2GM_1}{r}\;\; ,\quad \mathrm{for}\;\; 
r < r_0 \equiv r_{\mathrm{horizon}}= 2GM_1 \; .
\lab{Schwarzschild-metric-in}
\ee
\item
Reissner-Norstr\"{o}m black hole outside the horizon:
\be
A(r)\equiv A_{+}(r) = 1 - \frac{2GM_2}{r} + \frac{GQ^2}{r^2}\;\; ,
\quad \mathrm{for}\;\; r > r_0 \equiv r_{\mathrm{horizon}} \; ,
\lab{RN-metric-out}
\ee
where $Q^2 = 8\pi q^2 r_{\mathrm{horizon}}^4 \equiv 128\pi q^2 G^4 M_1^4$;
\end{itemize}
For the Maxwell subsystem we have $\cA_1 = \ldots =\cA_{D-1}=0$ everywhere and:
\begin{itemize}
\item
Coulomb field outside horizon:
\be
\cA_0 = \frac{\sqrt{2}\, q\, r_{\mathrm{horizon}}^2}{r} \;\; ,\quad \mathrm{for}\;\; 
r \geq r_0 \equiv r_{\mathrm{horizon}}  \; .
\lab{EM-out}
\ee
\item
No electric field inside horizon:
\be
\cA_0 = \sqrt{2}\, q\, r_{\mathrm{horizon}} = \mathrm{const} \;\; ,\quad \mathrm{for}\;\; 
r \leq r_0 \equiv r_{\mathrm{horizon}}  \; .
\lab{EM-in}
\ee
\end{itemize}

For the \textsl{WILL}-membrane subsystem the corresponding solution reads:
\be
X^0 \equiv t = \t \quad,\quad \th = \s^1 \quad,\quad \p = \s^2 \quad,\quad 
r (\t,\s^1,\s^2) = r_{\mathrm{horizon}} = \mathrm{const} \; ,
\lab{Schwarzschild-RN-sol}
\ee
where $A_{\pm}(r_{\mathrm{horizon}})=0$ , 
\textsl{i.e.}, the \textsl{WILL}-membrane ``sits'' on (materializes) the 
common event horizon of the pertinent black holes. Furthermore, the presence
of the \textsl{WILL}-membrane entails an important matching condition for the 
metric components along its surface\foot{The matching condition 
\rf{metric-match} corresponds to the statically soldering conditions in the 
phenomenological theory of
light-like thin shell dynamics in general relativity \ct{Barrabes-Israel}. } :
\be
\partder{}{r} A_{+}\bgv_{r=r_{\mathrm{horizon}}} -
\partder{}{r} A_{-}\bgv_{r=r_{\mathrm{horizon}}} = - 16\pi G \chi  \; ,
\lab{metric-match}
\ee
which yields the following relations between the parameters of the black
holes and the \textsl{WILL}-membrane ($q$ being its surface charge density) :
\be
M_2 = M_1 + 32\pi q^2 G^3 M_1^3
\lab{mass-match}
\ee
and for the brane tension $\chi$:
\be
\chi \equiv T_0 - 2q^2r_{\mathrm{horizon}} =q^2 G M_1 \quad, \;\;
\mathrm{i.e.} \;\; T_0 = 5 q^2 G M_1  \; .
\lab{tension-match}
\ee

Let us stress that the present {\em WILL}-brane models provide a systematic 
description of light-like branes from first principles starting with concise 
Weyl-conformal invariant actions
\rf{WI-brane}, \rf{E-M-WILL}--\rf{WILL-membrane+A-a}. As a consequence, these
actions also yield additional information impossible to obtain within the
phenomenological approach to light-like thin shell dynamics \ct{Barrabes-Israel}
(\textsl{i.e.}, where the membranes are introduced {\em ad hoc}),
such as the requirement that the light-like brane must sit on the (common) event
horizon(s) of the pertinent black hole(s). 
%%%%%%%%%%%%%%%%%%%%%%%%%%%%%%%%%%%%%%%%%%%%%%%%%%%%%%%%%%%%%%%%%%%%%%%%%%%%%%%%

\section{Conclusions and Outlook} 

In the present work we have discussed a novel class of Weyl-invariant $p$-brane 
theories whose dynamics significantly differs from ordinary Nambu-Goto $p$-brane
dynamics. The princial ingredients of our construction are:
\begin{itemize}
\item
Alternative non-Riemannian integration measure (volume-form) \rf{mod-measure-p}
on the $p$-brane world-volume independent of the intrinsic Riemannian metric;
\item
Acceptable dynamics in the novel class of brane models
(Eqs.\rf{WI-brane},\rf{WILL-membrane+A-a}) {\em naturally} requires the 
introduction of additional world-volume gauge fields.
\item
By employing square-root Yang-Mills actions for the pertinent
world-volume gauge fields one achieves manifest {\em Weyl-conformal
symmetry} in the new class of $p$-brane theories {\em for any $p$}.
\item
The brane tension is {\em not} a constant dimensionful scale given 
{\em ad hoc}, but rather it appears as a {\em composite} world-volume scalar 
field (Eqs.\rf{chi-2},\rf{chi-1},\rf{tension+A}) 
transforming non-trivially under Weyl-conformal transformations.
\item
The novel class of Weyl-invariant $p$-brane theories describes
intrinsically {\em light-like} $p$-branes for any even $p$ ({\em WILL}-branes).
\item
When put in a gravitational black hole background, the {\em WILL}-membrane
($p=2$) sits on (``materializes'') the event horizon.
\item
When moving in background product-spaces (``Kaluza-Klein'' context) the 
\textsl{WILL}-membrane describes {\em massless} modes, even though the
membrane is wrapping the extra dimensions and therefore aquiring non-trivial 
Kaluza-Klein charges.
\item
The coupled Einstein-Maxwell-{\em WILL}-membrane system \rf{E-M-WILL} possesses
self-consistent solution where the {\em WILL}-membrane serves as a 
material and electrically charged source for gravity and electromagnetism,
and it ``sits'' on (materializes) the common
event horizon for a Schwarzschild (in the interior) and Reissner-Nordstr\"{o}m
(in the exterior) black holes. Thus our model \rf{E-M-WILL} provides an
explicit dynamical realization of the so called ``membrane paradigm'' in the
physics of black holes \ct{membrane-paradigm}. 
\item
The {\em WILL}-branes could be good representations for the string-like objects
introduced by 't Hooft in ref.\ct{Hooft} to describe gravitational interactions
associated with black hole formation and evaporation, since as shown above the 
{\em WILL}-branes locate themselves automatically in the horizons and,
therefore, they could represent degrees of freedom associated particularly with 
horizons.
\end{itemize}

The novel class of Weyl-conformal invariant $p$-branes discussed above
suggests various physically interesting directions for further study:
quantization (Weyl-conformal anomaly and critical dimensions); supersymmetric
generalization; possible relevance for the open string dynamics (similar to the
role played by Dirichlet- ($Dp$-)branes); {\em WILL}-brane dynamics in more 
complicated gravitational black hole backgrounds (\textsl{e.g.}, Kerr-Newman).

\vspace{.1in}

%%%%%%%%%%%%%%%%%%%%%%%%%%%%%%%%%%%%%%%%%%%%%%%%%%%%%%%%%%%%%%%%%%%%%%%%%%%%%%%%
\textbf{Acknowledgements.}
{\small Two of us (E.G. and E.N.) are sincerely grateful to Plamen Fiziev
and the organizers of the \textsl{Second Workshop on Gravity, Astrophysics
and Strings} for the kind invitation to present there the above results. 
E.N. and S.P. are also thankful for hospitality and support to the
organizers of the 2nd Annual Meeting of the European
RTN \textsl{EUCLID}, Sozopol (Bulgaria), 2004. One of us (E.G.) thanks 
the Institute for Nuclear Research and Nuclear Energy (Sofia) and Trieste
University for hospitality. He also acknowledges useful conversations with
Gerard `t Hooft, Euro Spallucci and Stefano Ansoldi.

E.N. and S.P. are partially supported by Bulgarian NSF grants \textsl{F-904/99}
and \textsl{F-1412/04}.
Finally, all of us acknowledge support of our collaboration through the exchange
agreement between the Ben-Gurion Univesity of the Negev (Beer-Sheva, Israel) and
the Bulgarian Academy of Sciences.}
%%%%%%%%%%%%%%%%%%%%%%%%%%%%%%%%%%%%%%%%%%%%%%%%%%%%%%%%%%%%%%%%%%%%%%%%%%%%%%%%

%%%%%%%%%%%%%%%%%%%%%%%%%%%%%%%%%%%%%%%%%%%%%%%%%%%%%%%%%%%%%%%%%%%%%%%%%%%%%%%%
\end{document}